\documentstyle[12pt]{article}

\begin{document}

\begin{titlepage}
\begin{flushright}
\today \\
BA-TH/93-147\\
DPUR 65
\end{flushright}
\vspace{.5cm}
\begin{center}
{\LARGE A Coherent Understanding of Solvable Models

for Quantum Measurement Processes

} \vspace{1cm}

{\large Hiromichi NAKAZATO$^{(1)}$
 \& Saverio PASCAZIO$^{(2)}$ \\
   \quad \\
        $^{(1)}$Department of Physics, University of the Ryukyus,
  Okinawa 903-01, Japan \\
        $^{(2)}$Dipartimento di Fisica,
Universit\^^ {a} di Bari \\
and Istituto Nazionale di Fisica Nucleare, Sezione di Bari \\
I-70126  Bari, Italy

} \vspace{1cm}

{\small\bf Abstract}\\
\end{center}

{\small
By making use of Schwinger's oscillator model of angular momentum,
we put forward an interesting connection among three solvable Hamiltonians,
widely used for discussions on the
quantum measurement problem. This connection
implies that a particular macroscopic limit has to be taken for these models
to be physically sensible.}

\vspace*{.5cm}
\vfill

\end{titlepage}

     The quantum measurement problem \cite{von} \cite{Zurek} has been
a central
issue in the foundations and interpretation of quantum mechanics, and is
profoundly related to the investigation of the internal consistency of quantum
theory. In order to find a
resolution within the quantum mechanical framework, one usually
describes a quantum measurement process in the following way:
An elementary particle $Q$, on which the measurement is performed,
interacts with a detection system $D$, which is considered to be made
up of $N$ elementary constituents.  Generally, it is not easy to handle such a
complicated system like $Q+D$, however, it is also true that several solvable
models of measurement processes are now at hand. Among these are
the Cini \cite{Cini} and Coleman-Hepp \cite{AgBr} Hamiltonians and its
modified version \cite{NaPa3}, which were all proposed for consideration
of quantum
measurement processes, and the Jaynes-Cummings \cite{JC} model, that describes
the  interaction between a two-level atom and a single electromagnetic mode
in a cavity.

     Although these three solvable models were proposed independently,
a great deal of similarity exists among them. Indeed,
we can establish several links among them in certain limits \cite{NaPa1}
\cite{NaPa3}. This is both interesting and rather curious.
The purpose of this letter is
to gain a more comprehensive understanding on such
models for quantum measurement processes, and in particular on
their macroscopic limits, by making use
of Schwinger's double-oscillator method \cite{Schwinger}
\cite{Sakurai}.
We stress that in all the above-mentioned cases, both the macroscopic system
and the measurement process are described according to the quantum mechanical
laws, and no ^^ ^^ classical" behavior is postulated.

     Let us first briefly review Schwinger's method.
Schwinger's construction of the algebra of angular momentum
originates from two independent harmonic oscillators,
henceforth conventionally denoted as type $+$ and $-$.
The creation and annihilation operators
$a^{\dagger}_{\pm},  a_{\pm}$  obey the standard boson commutation relations:
\begin{eqnarray}
[a_\pm,a^{\dagger}_\pm] & = & 1 ,   \nonumber \\
 \left[ {\cal N}_\pm,a_\pm \right] & = & - a_\pm ,   \label{eq:standboson} \\
 \left[ {\cal N}_\pm,a^\dagger_\pm \right] & = & a^\dagger_\pm , \nonumber
\end{eqnarray}
where ${\cal N}_\pm$ are the number operators.
All commutators of operators of different oscillators
are assumed to vanish:
This is what is meant by {\em independent} oscillators.

One can construct the eigenstates of ${\cal N}_\pm$ by acting with
$a^{\dagger}_{\pm}$ on the {\em vacuum} state, defined by
\begin{equation}
a_\pm \vert0,0> = 0 .
\label{eq:vacdef}
\end{equation}
For instance, one has
\begin{equation}
\vert n_+,n_-> = \frac{1}{\sqrt{n_+!}} \frac{1}{\sqrt{n_-!}}
(a^\dagger_+)^{n_+} (a^\dagger_-)^{n_-} \vert 0,0> .
       \label{eq:numstate}
\end{equation}

Schwinger's interesting and brilliant idea was to define [we set
$\hbar =1$ in eqs.(\ref{eq:brilly1})-(\ref{eq:angstate})]
\begin{eqnarray}
J_\pm & \equiv &  a^{\dagger}_\pm a_\mp ,       \label{eq:brilly1} \\
J_3 & \equiv & \frac{1}{2} \left( a^{\dagger}_+ a_+
             - a^{\dagger}_- a_-  \right) .
     \label{eq:brilly2}
\end{eqnarray}
The above operators obey the standard commutation
relations for angular momentum:
\begin{eqnarray}
[J_3,J_\pm] & = & \pm J_\pm ,   \nonumber \\
\left[ J_+,J_- \right] & = & 2 J_3 .
\label{eq:standang}
\end{eqnarray}
Moreover
\begin{equation}
{\bf J}^2 = J_3^2 + \frac{1}{2} \left( J_+J_- + J_-J_+  \right)
    = \frac{{\cal N}_{\rm tot}}{2} \left( \frac{{\cal N}_{\rm tot}}{2}
      + 1 \right),
       \label{eq:Casimir}
\end{equation}
where ${\cal N}_{\rm tot} \equiv {\cal N}_+ + {\cal N}_-$.
By identifying
\begin{eqnarray}
n_+ \leftrightarrow  j+m , & &  \qquad
          \frac{n_+ + n_-}{2} \leftrightarrow  j ,  \nonumber \\
n_- \leftrightarrow  j-m , & &  \qquad
          \frac{n_+ - n_-}{2} \leftrightarrow  m ,
\label{eq:idntf}
\end{eqnarray}
where $j,m$ are the eigenvalues of ${\bf J}^2$ and $J_3$, respectively,
one can construct the familiar simultaneous eigenstates of
${\bf J}^2$ and $J_3$:
\begin{equation}
\vert j,m> \equiv \frac{1}{\sqrt{(j+m)!}} \frac{1}{\sqrt{(j-m)!}}
(a^\dagger_+)^{j+m} (a^\dagger_-)^{j-m} \vert 0> .       \label{eq:angstate}
\end{equation}

Let us now apply Schwinger's procedures to show how it is possible
to obtain the Coleman-Hepp Hamiltonian \cite{AgBr} from Cini's \cite{Cini}.
This will have interesting spin-offs: The conditions under which the two
Hamiltonians can be identified 
\cite{NaPa1} will suggest a link with the famous
Jaynes-Cummings Hamiltonian \cite{JC}, widely used in laser theory
and recently utilized to discuss fundamental issues in quantum mechanics
\cite{NaPa3} \cite{JCderiv}.
Here and in the following we shall restrict our attention to the interaction
term of the total Hamiltonian, and shall comment on the role of the
free part whenever necessary.

In the model proposed by Cini, a quantum particle interacts with a
detector $D$, made up of $N$ particles, each of
which is assumed to have only two possible states, say a
ground state $\omega_{-}$ and an excited or ionized state
$\omega_{+}$.
The interaction Hamiltonian is taken to be
\begin{equation}
H_{\rm C} = g \left( a^\dagger_- a_+ + a^\dagger_+a_- \right) ,
\label{eq:HCini}
\end{equation}
where $g$ is a coupling constant and
$a^\dagger_\pm,  a_\pm$  are
the creation and annihilation operators for the states
$\omega_\pm$.
We are here neglecting a (trivial)
$\frac{1}{2} (1+\tau_{3})$ factor,
whose only effect is to select
which of the two states of the Q particle interacts with the
detector.

By making use of eq.(\ref{eq:brilly1}) we can write
\begin{equation}
H_{\rm C} = g \left( J_+ + J_- \right) = g J_1 .
\label{eq:reCini}
\end{equation}
Notice that $H_{\rm C}$ is essentially expressed as one of the group elements
of angular momentum.

The Coleman-Hepp (CH) or AgBr Hamiltonian \cite{AgBr} \cite{AgBrmore}
describes the interaction between a
particle $Q$ and a one-dimensional $N$-spin array
($D$-system).
One can think, for instance, of a linear emulsion  of AgBr
molecules, the {\em down} state  corresponding  to  the  undivided
molecule, and the {\em up} state corresponding to the  dissociated
molecule (Ag and Br atoms).
The particle and each molecule interact via a spin-flipping local potential,
according to the following interaction Hamiltonian
\begin{equation}
H_{\rm CH} = \sum_{n=1}^{N}  V(\widehat{x}- x_n)  \sigma_{1}^{(n)} \ ,
\label{eq:CHorig}
\end{equation}
where $\widehat{x}$ is the position of
the particle, $V$ is a real  potential,  $x_n \; (n=1,...,N)$
are the positions of the scatterers in the  array
and $\sigma_{1}^{(n)}$ is the Pauli matrix acting on the $n$th site.
This Hamiltonian is a nice model of a typical measurement process and can
be solved exactly if the $Q$ particle is ^^ ^^ ultrarelativistic",
namely if its free Hamiltonian is written as
$H_{Q}= c\widehat{p}$, $\widehat{p}$ being the particle's momentum.
In such a case, the $S$-matrix is readily computed as
\begin{equation}
S^{[N]} = \exp \left( -i\frac{V_{0} \delta}{\hbar c}
\sum_{n=1}^{N} \sigma_1^{(n)} \right)
   =  \exp \left( -i\frac{V_{0} \delta }{\hbar c}
 N \Sigma_1^{(N)} \right) ,
\label{eq:Smatr}
\end{equation}
where $V_{0} \delta \equiv \int_{-\infty}^{\infty} V(x)dx$
is the integrated strength of the potential,
$q \equiv \sin^2 (V_{0} \delta / \hbar c )$ is viewed as the
probability of dissociating one AgBr molecule,
and we defined the average spin
\begin{equation}
\Sigma^{(N)}_{j}  =  \frac{1}{N}
 \sum_{n=1}^{N}
\sigma_{j}^{(n)}, \qquad  j=1,2,3.      \label{eq:Sigmaj}
\end{equation}
If the initial $D$-state is taken to be the ground state
$\vert  0>_N$ ($N$ spins down), and the initial $Q$-state is a plane wave,
the evolution is
\begin{equation}
S^{[N]} \vert p, 0>_N
   = \sum_{j=0}^N {N\choose j}^{1/2}
  \left( -i\sqrt{q} \, \right)^j \left( \sqrt{1-q} \, \right)^{N-j}
 \vert p, j>_N,
\label{eq:evSvac}
\end{equation}
where $\vert j>_N$ is the (symmetric) state with $j$
dissociated molecules,
and we have used the notation $\vert p,j>_N = \vert p>\vert j>_N$.
The right hand side in eq.(\ref{eq:evSvac}) is a generalized
[SU(2)] coherent state. Notice the (square root of the) binomial
distribution in the above equation.

Observe that
the operators $N \Sigma^{(N)}_j$ form a unitary representation
of SU(2), and therefore satisfy the commutation relations (\ref{eq:standang}).
Notice that in this representation, the Cini Hamiltonian
(\ref{eq:reCini}) is expressed as
\begin{equation}
H_{\rm C} = g \left( N \Sigma_+^{(N)} + N \Sigma_-^{(N)} \right) =
   g N \Sigma_1^{(N)} = g \sum_{n=1}^{N} \sigma_1^{(n)} .
\label{eq:re2Cini}
\end{equation}
This expression is very similar to the CH Hamiltonian (\ref{eq:CHorig}),
and indeed, $H_{\rm C}$ appears as a particular case of $H_{\rm CH}$:
Consider either the situation in which the
spins are all placed at the same position
\begin{equation}
x_n \equiv x_0 = 0, \qquad \forall n=1, \ldots , N
\label{eq:sameplace}
\end{equation}
or the (similar) situation of an ^^ ^^ average" potential
over the positions of the scatterers $x_n$ (replace
$V(\widehat{x}-x_n)$
with its average, say $\overline{V} (\widehat{x})$, and call the latter
$V(\widehat{x})$ again.
Both cases have been analyzed in Ref.~\cite{NaPa1}):
$H_{\rm C}$ and $H_{\rm CH}$ coincide if one
takes $V(\widehat{x}) =$ constant $=g$.
It is necessary to observe, in this context,
that unlike CH, Cini's Hamiltonian does not contain a free part,
involving the $Q$-particle coordinates, so that one is forced to assume
that the $Q$ and $D$ systems are in contact for a certain
^^ ^^ interaction time" $\tau$.
In this sense, the CH Hamiltonian appears more self-contained,
because the interaction time is naturally given by
$\tau = L/c$, where $L$ is the total length of the spin array and $c$
the $Q$-particle speed.
Interestingly, we shall also see that $\tau$ cannot be arbitrarily
chosen for $H_{\rm C}$: This will shed light on Cini's
choice for $\tau$.

In order to establish a link with the Jaynes-Cummings model,
we need to consider the weak-coupling, $N \rightarrow \infty$
limit of the CH Hamiltonian \cite{NaPa3}.
Notice that $H_{\rm CH}$ is invariant under exchange of
spins (molecules) in
the array. Therefore, if we call ${\cal P}_N$ the group of permutations on
$\{ 1, \ldots, N \}$, we can restrict our attention to the
${\cal P}_N$-invariant
sector ${\cal H}_N$ of the bigger Hilbert space ${\cal H}_{\{N\}}$ of the
$N$ spins. In the following, we shall concentrate
our analysis on the {\em symmetrized} case.

The weak-coupling, $N \rightarrow \infty$ limit of $H_{\rm CH}$ can be
consistently taken only if we consider a modified version of
the CH model, that enables us
to take into account the possibility of energy
exchange between the particle and the spin system.
One writes the interaction part as
\begin{eqnarray}
H'_{\rm CH} & = & \sum_{n=1}^{N}  V(\widehat{x}- x_n)
  \sigma_{1}^{(n)} \exp \left( i \frac{\omega}{c}
  \sigma_{3}^{(n)} \widehat{x} \right)  \nonumber \\
    & = & \sum_{n=1}^{N} V(\widehat{x}- x_n)
  \left[ \sigma_{+}^{(n)} \exp \left( -i \frac{\omega}{c}
  \widehat{x} \right) + \sigma_{-}^{(n)} \exp \left( + i \frac{\omega}{c}
  \widehat{x} \right) \right] , \label{eq:Hmod}
\end{eqnarray}
and adds the free Hamiltonian of the spin array
$H_{D} = \frac{1}{2} \hbar \omega
\sum_{n=1}^{N} \left( 1+\sigma_{3}^{(n)} \right)$.
The reasons for this modification and additional
details can be found in Ref.~\cite{NaPa3}:
It is remarkable that the model remains solvable
provided that a ^^ ^^ resonance condition" (implicitly assumed in
writing the above expressions) is met, because in this way the energy
acquired or lost by the $Q$-particle in every single interaction
matches exactly the energy gap between the two spin states.
In fact, there is a deep reason for this: It is practically equivalent
to the rotating-wave approximation in optics.

The weak-coupling, $N \rightarrow \infty$ limit of $H_{\rm CH}$ can be
consistently taken by performing a contraction on SU(2)~\cite{contrac}:
We start from a linear transformation of the generators of SU(2)
\begin{equation}
\left( \begin{array}{c}
h_+ \\
h_- \\
h_3 \\
1
\end{array} \right) =
\left( \begin{array}{cclc}
N^{-1/2} & & & \\
 & N^{-1/2} & & \\
 & & 1 & N/2 \\
 & & & 1
\end{array} \right)
\left( \begin{array}{c}
N \Sigma^{(N)}_+  \\
N \Sigma^{(N)}_-  \\
N \Sigma^{(N)}_3/2  \\
{\bf 1}^{(N)}
\end{array} \right) ,
         \label{eq:newbasis}
\end{equation}
so that the commutation properties for ${\bf h}, 1$ are
\begin{eqnarray}
\left[ h_3 , h_\pm \right] & = & \pm h_\pm ,
  \nonumber \\
\left[ h_- , h_+ \right] & = & 1 - \frac{2}{N} h_3 ,
  \nonumber \\
\left[ {\bf h} , 1 \right] & = & 0 ,
         \label{eq:newcomm}
\end{eqnarray}
and yield, in the $N \to \infty$ limit,
the standard boson commutation relations:
\begin{eqnarray}
h_+ = \frac{1}{\sqrt{N}} \sum_{n=1}^{N} \sigma_{+}^{(n)}
 = \sqrt{N} \Sigma_{+}^{(N)}
   & \stackrel{N \rightarrow \infty}{\longrightarrow} & a^\dagger, \nonumber \\
h_- = \frac{1}{\sqrt{N}} \sum_{n=1}^{N} \sigma_{-}^{(n)}
 = \sqrt{N} \Sigma_{-}^{(N)}
& \stackrel{N \rightarrow \infty}{\longrightarrow} & a,
 \label{eq:sumup22}    \\
h_3 = \frac{1}{2} \sum_{n=1}^{N} \left( 1 + \sigma_{3}^{(n)} \right)
 = \frac{N}{2} \left( {\bf 1}^{(N)} + \Sigma_{3}^{(N)} \right)
   & \stackrel{N \rightarrow \infty}{\longrightarrow} & {\cal N}
   \equiv a^\dagger a .
    \nonumber
\end{eqnarray}
Now assume again, for simplicity, that all spins are
placed at the same position, so that the Hamiltonian (\ref{eq:Hmod})
becomes
\begin{eqnarray}
H'_{\rm CH} \! \!
   & = & V(\widehat{x}) \sum_{n=1}^{N}
  \left[ \sigma_{+}^{(n)} \exp \left( -i \frac{\omega}{c}
  \widehat{x} \right) + \sigma_{-}^{(n)} \exp \left( + i \frac{\omega}{c}
  \widehat{x} \right) \right]  \nonumber \\
   & = & V(\widehat{x}) \sqrt{N} \left[
         \exp \left( -i \frac{\omega}{c} \widehat{x} \right)
         \frac{1}{\sqrt{N}} \sum_{n=1}^{N} \sigma_{+}^{(n)}
         + \exp \left( + i \frac{\omega}{c} \widehat{x} \right)
         \frac{1}{\sqrt{N}} \sum_{n=1}^{N} \sigma_{-}^{(n)} \right]
\nonumber  \\
   & \stackrel{N \rightarrow \infty, \, V\sqrt{N}=u}{\longrightarrow}
   & u(\widehat{x})
         \left[ a^\dagger \exp \left( -i \frac{\omega}{c} \widehat{x} \right)
         + a \exp \left( i \frac{\omega}{c} \widehat{x} \right)
         \right] = H_{\rm JC} .
  \label{eq:reH}
\end{eqnarray}
The connection with a ^^ ^^ maser" system is manifest: In the
weak-coupling, $N \rightarrow \infty$ limit we obtain
the Jaynes-Cummings (JC) \cite{JC} Hamiltonian.
(Strictly speaking, the JC Hamiltonian differs from the above one
because it contains terms of the type $\tau_\pm$, instead of
$\exp \left(\pm i \frac{\omega}{c} \widehat{x} \right)$, $\tau_\pm$ being the
raising/lowering operators for a two-level system.
In the case we are considering the $Q$ particle has a continuous spectrum,
and can exchange an arbitrary number of quanta of energy $\hbar \omega$.
Clearly, this difference is not important for our analysis.)

It is necessary to observe that the above limit is taken by keeping
$V(\widehat{x}) \sqrt{N} =
u(\widehat{x})$ finite, while sending $N$ to infinity.
This is just a {\em consequence} of the (physically appealing)
requirement that the average number of dissociated
molecules, namely the quantity $qN$, with
$\sqrt{q} \simeq V_0 \delta / \hbar c = \int V(y) dy / \hbar c$
[see the definition after eq.(\ref{eq:Smatr})],
is kept finite in the $N \rightarrow \infty$ limit.

Incidentally, notice that in this limit the
free Hamiltonian of the spin array, introduced after eq.(\ref{eq:Hmod}),
yields just the free Hamiltonian of a single-mode electromagnetic cavity:
\begin{equation}
H_{D} = \frac{1}{2} \hbar \omega
\sum_{n=1}^{N} \left( 1+\sigma_{3}^{(n)} \right)
  \stackrel{N \rightarrow \infty}{\longrightarrow}
   \hbar \omega {\cal N} = \hbar \omega a^\dagger a = H_{D}^{\rm JC} .
          \label{eq:JCfree}
\end{equation}
The $N$-spin system behaves, in the above-mentioned limit,
as a ^^ ^^ cavity", in which boson-like excitations (collective modes)
can be created, as a consequence of the interaction with the $Q$-particle.

The $S$-matrix of the JC Hamiltonian is
\begin{equation}
S = \exp \left[ -i\frac{u_{0} \delta}{\hbar c}
         \left( a^\dagger \exp \left[ -i \frac{\omega}{c} \widehat{x} \right]
         + a \exp \left[ i \frac{\omega}{c} \widehat{x} \right]
                       \right) \right],
  \label{eq:SJC}
\end{equation}
where $\int u(x) dx = u_0 \delta$,
and we assumed, for simplicity and without loss of generality,
that $\delta$ is the same quantity used in eq.(\ref{eq:Smatr}).
If we take the initial state to be $\vert p,0>=\vert p>\vert 0>$,
where $\vert 0>$ is the ground state of the maser cavity, the evolution is
\begin{equation}
S \vert p, 0> = e^{- \overline{\kappa}/2}
   \sum_{j=0}^\infty \frac{(-i \sqrt{\overline{\kappa}} \, )^j}{\sqrt{j!}}
   \vert p - j \frac{\hbar \omega}{c}, j>,
  \qquad \overline{\kappa} = \left( \frac{u_0 \delta}{\hbar c} \right)^2 ,
\label{eq:SvacJC}
\end{equation}
where $\vert j>$ is the number state of the cavity
and $\overline{\kappa}$ is the average number of boson excitations in
the cavity. This reduces to the
$N \rightarrow \infty$, $qN \equiv \overline{\kappa}$ finite
limit of eq.(\ref{eq:evSvac}) when we neglect the
energy differences between spin
states, i.e., $\omega=0$.
[The general case, with $\omega \neq 0$, has been analyzed in
Ref.~\cite{NaPa3} and yields a final state that is essentially and conceptually
similar to (\ref{eq:evSvac}).]

We wish to stress that the connection between
$H'_{\rm CH}$ and $H_{\rm JC}$ has been proven under the condition that all
spin were at the same position in space, like in eq.(\ref{eq:sameplace}).
The fact that such a condition is unnecessary was already stressed in
Ref.~\cite{NaPa3}, and in fact a rigorous derivation, that does not make use of
(\ref{eq:sameplace}) has recently been found \cite{rigor}.

We emphasized before
that the $N \rightarrow \infty$ limit was taken by keeping
$\sqrt{qN} = (V_0 \delta / \hbar c) \sqrt{N}
= (\int V(y) dy / \hbar c)\sqrt{N}$ finite.
Since the $Q+D$ system is isolated, its total energy is conserved,
so that $qN\hbar \omega$, the energy necessary to provoke the dissociation of
$qN$ AgBr molecules (namely, the average
energy ^^ ^^ stored in the detector"),
is also the energy lost by the $Q$ particle. Such an energy {\em must} be
finite, because if it were not finite we would encounter a pathological
situation of the model, in which the $Q$ particle has a negative energy:
The above limit
appears therefore as a {\em forced choice from the physical point of view}.
Notice also that, in the above limit,
the binomial statistics in eq.(\ref{eq:evSvac}) yields
the Poisson statistics of eq.(\ref{eq:SvacJC}). Obviously, this could
be accomplished {\em only} by keeping $qN = \overline{\kappa}$
finite.
Returning to our AgBr model, the probability $q$ of one
spin flip or one AgBr molecule dissociation becomes very low,
of order O($N^{-1}$):
In the
macroscopic limit, the event of one dissociation becomes therefore a
{\em rare event} in the statistical sense, for the AgBr model to
make sense from a physical point of view.

In this context, observe that Cini redefined his coupling constant as
\begin{equation}
g \equiv \frac{g_0}{\sqrt{N}},       \label{eq:g0}
\end{equation}
so that
$\tau_{0} = \hbar/g  \sqrt{N}  =  \hbar/g_{0}$,  the  time required
to ionize the first particle of the initially  discharged
detector, becomes independent of $N$, the number of elementary
constituents of his ^^ ^^ detector".
Since $g$ is nothing but our $V_0 \delta \propto \sqrt{q}$,
Cini's intuition appears as a {\em straight consequence of the
macroscopicity of the detection system}.
Indeed, according to the above discussion, one must require that
$g^2 \propto q =$O($N^{-1}$), for the model to be physically sensible.

There are also profound links between the limiting procedure considered in this
letter and van Hove's ^^ ^^ $\lambda^2 T$" limit \cite{vanHove}.
This is easily evinced from the following example: Set $x_n \equiv na$
in eq.(\ref{eq:CHorig}) or (\ref{eq:Hmod}) (constant spacing
between adjacent potentials).
The free part of the JC Hamiltonian is
$H_{Q}= c\widehat{p}$, so that the particle travels with constant speed
$c$, and interacts with the detector for a time
$T = Na/c$, where $Na$ is the total length of the detector.
Since the coupling constant $\lambda \equiv g \propto V_0 \delta$,
one gets
$\lambda^2 T = g^2 Na/c \propto qN$. (Observe that the
^^ ^^ lattice spacing" $a$, the inverse of which corresponds to a density
in our 1-dimensional
model, is kept constant in the above limit.)
Note also that the hypothesis of photon-like dispersion-free $Q$ particles
($H_{Q}= c\widehat{p}$), typical of CH-like models,
does not appear very restrictive: In similar cases,
such as the JC model \cite{JCderiv}, in order to solve the equations
of motion, it is often assumed that the particle has constant speed while
travelling through the maser. Incidentally, we stress that this assumption
is in line with our $qN$-finite ansatz: The energy lost by the particle
during the interaction is negligible compared to
its kinetic energy ($qN \hbar \omega \ll cp$).

We have mentioned some aspects of the quantum theory of
measurement and stressed the importance of performing a full
quantum mechanical analysis of the (macroscopic)
^^ ^^ detector".
The purpose of this work was to clarify the similarity among some model
detectors, and illustrate that
no ^^ ^^ classical" behavior is to be postulated.

Many important topics, such as the problem of the
loss of quantum coherence \cite{MHS} and the general argument of
^^ ^^ objectification" \cite{Busch}, namely the emergence of definite
outcomes in a measurement process, have not been addressed.
The problem of a consistent description of the interaction
between a macroscopic and a microscopic object \cite{Haake}
is delicate and, in our opinion, still open.
The unitarity of the quantum evolutions appears to many physicists
as an insuperable problem, and has led some authors
to conjecture that nonlinear terms should be added
to the Schr\"odinger equation \cite{nonlin}.
This is a logically consistent
possibility that requires further investigation.

At the present stage, we feel that no solution, among those proposed,
is fully satisfactory. Nevertheless, we are confident that further progress on
these foundamental quantum mechanical problems will also come from
^^ ^^ technical" analyses. Solvable models often provide excellent clues
in this direction.
\vspace*{.3cm}

The authors acknowledge interesting remarks by Professors Y.G.~Lu and
L.~Accardi.  They thank the University of the Ryukyus Foundation and
the Italian National Institute for Nuclear Physics (INFN) for the financial
support. This work was also partially supported by the Grant-in-Aid
for Scientific Research of the Ministry of Education, Science and Culture,
Japan (03854017) and by Italian Research National Council
(CNR) under the bilateral
projects Italy-Japan n.91.00184.CT02 and n.92.00956.CT02.



\end{document}